\documentclass[aps,prd,reprint,preprintnumbers,superscriptaddress,showpacs,twocolumn]{revtex4-1}
\usepackage{latexsym,graphicx,amssymb,amsmath,mathrsfs}
\usepackage{setspace,bm}
\usepackage[breaklinks, colorlinks=true, pdfstartview=FitV, linkcolor=red, citecolor=blue, urlcolor=blue]{hyperref}
\usepackage[usenames]{color}
\usepackage{latexsym}
\usepackage{epstopdf}
\usepackage{mathtools}
\usepackage{url}
\usepackage{comment}
\usepackage{braket}
\usepackage{CJKutf8}

\begin{document}

\title{The canonical approach at high temperature revisited}

\author{Kouji Kashiwa}
\email[]{kashiwa@fit.ac.jp}
\affiliation{Department of Computer Science and Engineering, Faculty of Information Engineering, Fukuoka Institute of Technology, Fukuoka 811-0295, Japan}

\author{Hiroaki Kouno}
\affiliation{Department of Physics, Saga University, Saga 840-8502, Japan}

\begin{abstract}
This paper discusses a paradox encountered when employing the canonical approach, particularly in the high-temperature region where the Roberge-Weiss transition exists at finite imaginary chemical potential.
The paradox is that the results obtained using the canonical approach cannot match the correct results in that region.
We show that the paradox originates from the Roberge-Weiss transition in the infinite-size system, which is linked to the non-trivial Polyakov-loop sectors.
Furthermore, it is shown that this paradox disappears in finite-size systems because of the smearing effect for the Roberge-Weiss transition, which validates the use of the canonical approach in lattice QCD simulations.
\end{abstract}
\maketitle

\section{Introduction}

Exploring the non-perturbative properties of quantum chromodynamics (QCD) at finite temperature (T) and finite real quark chemical potential ($\mu_\mathrm{R}$) is an interesting and crucial study for understanding our universe; see Ref.\,\cite{Fukushima:2010bq} as an example.
At finite $T$ but $\mu_\mathrm{R}=0$, the lattice QCD simulation is feasible; thus, non-perturbative properties can be investigated exactly using the simulation in principle.
However, at finite $\mu_\mathrm{R}$, QCD suffers from the sign problem.
Consequently, several methods are required to control the problem; see Ref.\,\cite{deForcrand:2010ys,Nagata:2021ugx} as an example.

To control the sign problem, some methods have been proposed so far; the famous examples are the reweighting method~\cite{Ferrenberg:1988yz,Fodor:2001au}, the Taylor expansion method~\cite{Allton:2002zi,Gavai:2003mf}, the canonical approach~\cite{Roberge:1986mm,Barbour:1988ax,Hasenfratz:1991ax}, and the analytic continuation method~\cite{deForcrand:2002hgr}.
Unfortunately, these methods are restricted in the $\mu_\mathrm{R}/T < 1$ region.
To go beyond restrictions, some other approaches have recently attracted much more attention, such as the Lefschetz thimble method~\cite{Witten:2010cx,Cristoforetti:2012su,Fujii:2013sra}, the complex Langevin method~\cite{Klauder:1983sp,Parisi:1984cs}, and the path optimization method manifold~\cite{Mori:2017nwj,Alexandru:2018fqp,Bursa:2018ykf}, but they are still far from perfection; these are based on the complexified dynamical variable approach.
In this work, we focus on the canonical approach.

It is known that the imaginary chemical potential ($\mu_\mathrm{I}$) region has a direct relation to the realistic real chemical potential region; for example, see Ref.\,\cite{deForcrand:2010ys,Kashiwa:2019ihm}.
That is, we can construct the grand canonical partition function at finite $\mu=(\mu_\mathrm{R},0)$ based on that at finite $\mu=(0,\mu_\mathrm{I})$~\cite{Roberge:1986mm}.
This is the so called canonical approach~\cite{Alexandru:2005ix,deForcrand:2006ec,Fukuda:2015mva,Bornyakov:2016wld,Oka:2017kny,Wakayama:2018wkc}.
This means that we can construct the canonical partition function with a fixed quark number using the grand canonical partition function at finite $\mu_\mathrm{I}$.
This indicates that we can examine the validity of the analytical method via the imaginary chemical potential region. 
In this paper, we discuss the paradox in the canonical approach at high temperature.

At finite $\mu_\mathrm{I}$, there are special properties such as Roberge-Weiss (RW) periodicity, the RW transition, and the RW endpoint:
\begin{description}
    \item[RW periodicity]
    At finite $\mu_\mathrm{I}$, thermodynamic quantities and order parameters exhibit nontrivial $2\pi/3$ periodicity along the $\theta$-axis where $\theta \equiv \mu_\mathrm{I}/T$~\cite{Roberge:1986mm}.
    This periodicity is usually called RW periodicity.
    \item[RW transition]
    The origin of the RW periodicity differs in the low- and high-temperature regions.
    This difference arises from the balance between gluon and quark contributions in the grand canonical partition function (${\cal Z}_\mathrm{GC}$).
    At $\theta=(2k-1)\pi/3$ with $k \in \mathbb{Z}$, several quantities exhibit singularities at high $T$~\cite{Roberge:1986mm}.
    The odd and even quantities against $\theta$ have first-order and second-order singularities, respectively. 
    Such singularity characterizes the phase transition, which is called the RW transition.
    \item[RW endpoint]
    Several quantities are smoothly oscillating along the $\theta$-axis at low $T$.
    However, these exhibit singularities at high $T$, as explained above.
    Due to this difference, there should be an endpoint of the first-order RW transition line.
    This endpoint is called the RW endpoint.
    Also, the temperature of the RW endpoint is referred to as the RW endpoint temperature $T_\mathrm{RW}$.
    The order of the endpoint depends on the current mass of the quark; see Ref.\,\cite{D'Elia:2009qz,Bonati:2010gi,Kashiwa:2013rm,Bonati:2014kpa} as an example.
    \item[$\mathbb{Z}_3$ images]
    Regions divided by the RW transition at high $T$ are characterized by the $\mathbb{Z}_3$ images which are characterized by the one trivial Polyakov-loop sector (phase is about $0$) and the two non-trivial Polyakov-loop sectors (phases are about $2\pi/3$ and $4\pi/3$).
\end{description}
For more details on the imaginary chemical potential, see Ref.\,\cite{Kashiwa:2019ihm} as an example.

The canonical approach, which is based on the Fourier transformation and the fugacity expansion, has been used to investigate the $\mu_\mathrm{R}$ region with lattice QCD simulations and QCD effective models.
In the canonical approach, there is a strong impact from the RW periodicity, as explained in Sec.\,\ref{Sec:Canonical_method}.
In this study, we discuss the paradox that the canonical method may deviate from the correct results in the high $T$ region, where $T$ is higher than the RW transition temperature.
It should be noted that the high-temperature region has recently attracted much more attention because some interesting insights are newly proposed, such as some exotic phases; for example, see Refs.\,\cite{Fujimoto:2025sxx,Cohen:2023hbq}. 
This is also the related motivation of this study.

This paper is organized as follows.
In the next section, we explain the properties of QCD in the $\mu_\mathrm{I}$ region.
Section\,\ref{Sec:Canonical_method} explains the canonical approach and its paradox.
Section\,\ref{sec:Discussions} presents discussions on the paradox.
Section\,\ref{sec:Summary} is devoted to the summary of this research.

\section{Canonical approach}
\label{Sec:Canonical_method}

It is well known that the canonical partition function (${\cal Z}_\mathrm{C}$) with a fixed quark number ($Q$) can be constructed using the grand canonical partition function (${\cal Z}_\mathrm{GC}$) at finite $\theta$.
The relationship is given as
\begin{align}
    {\cal Z}_\mathrm{C} (Q)
    &= \sum_{n=-\infty}^\infty \bra{n} e^{-\beta \hat{\cal H}} \delta (\hat{N}-Q) \ket{n} 
\nonumber\\
    &= \frac{1}{2\pi} \int_{-\pi}^\pi
       e^{i Q \theta} {\cal Z}_\mathrm{GC} (\theta) \, d\theta,
\label{Eq:canonical}
\end{align}
where $\hat{\cal H}$ denotes the Hamiltonian, $n \in \mathbb{Z}$ represents the eigenvalues of the quark number operator $\hat{N}$ and $Q \in \mathbb{Z}$.
This is based on the Fourier series of the delta function.
This formulation has already been applied to both lattice QCD simulations~\cite{Alexandru:2005ix,deForcrand:2006ec,Fukuda:2015mva,Bornyakov:2016wld,Oka:2017kny,Wakayama:2018wkc} and QCD effective model calculations~\cite{Morita:2015tma,Wakayama:2020dzz,Kashiwa:2021czl,Kashiwa:2021til,Kashiwa:2021toz}; see Ref.\,\cite{Adam:2026rnt} as an example of recent progress.
As expected, the canonical partition function obeys the canonical ensemble, allowing the evaluation of several physical quantities.

With the fugacity expansion, we obtain the grand canonical partition function at finite $\mu_\mathrm{R}$ in terms of the canonical partition functions as
\begin{align}
    {\cal Z}_\mathrm{GC}(\mu_\mathrm{R})
    &= \sum_{Q=-\infty}^\infty
       \exp \Bigl( {Q \frac{\mu_\mathrm{R}}{T}} \Bigr)
       {\cal Z}_\mathrm{C}(Q)
       \nonumber\\
    &= \sum_{Q=-\infty}^\infty
       \xi^Q
       {\cal Z}_\mathrm{C}(Q),
\label{Eq:fugacity}
\end{align}
where $\xi$ is the so called fugacity defined as
\begin{align}
    \xi &= \exp \Bigl( {Q \frac{\mu_\mathrm{R}}{T}} \Bigr).
\end{align}
In the canonical approach, we usually discuss the grand canonical partition function and related quantities, but the canonical partition function itself is also interesting. 
Actually, the canonical partition functions relate to the multiplicity distribution that can be extracted from collision experiments; see Refs.\,\cite{STAR:2010mib,Luo:2012kja} for the experimental data and the recent review~\cite{Fukushima:2020yzx}.
Actually, this fact was used to combine lattice QCD data and experimental data at finite density through Lee-Yang zeros~\cite{yang1952statistical,lee1952statistical}; see Ref.\,\cite{Nakamura:2013ska} as an example.

The canonical partition function (\ref{Eq:canonical}) with a fixed quark number $Q$ can be constructed by using ${\cal Z}_\mathrm{GC}$ at finite $T$ and $\theta$ as
\begin{align}
    &{\cal Z}_\mathrm{C} (Q)
    \nonumber\\
    &= \frac{1 + z^{Q} + z^{2Q}}{2\pi}
       \int_{-\pi/3}^{\pi/3}
       e^{i Q \theta} {\cal Z}_\mathrm{GC} (\theta) \, d\theta
       \nonumber\\
    &=
    \begin{dcases}
    \frac{3}{2\pi}  
    \int_{-\pi/3}^{\pi/3}
       e^{i Q \theta} {\cal Z}_\mathrm{GC} (\theta) \, d\theta
       & (Q= 3 k)\\
       ~0 & (Q \neq 3 k)
    \end{dcases}
    ,
\label{Eq:canonical_RW_p}
\end{align}
where $k \in \mathbb{Z}$ and
\begin{align}
    z = \exp \Bigl( \frac{2\pi i}{3} \Bigr),
\end{align}
is the ${\mathbb Z}_3$ factor.
In this study, we consider the three color system, and thus the center symmetry is $\mathbb{Z}_3$. 
The above property (\ref{Eq:canonical_RW_p}) is also valid for almost all quantities except the Polyakov-loop; the problem for the Polyakov loop is known as the Polyakov-loop paradox~\cite{Kratochvila:2006jx}, and is resolved by using the modified Polyakov loop in Ref.\,\cite{Kashiwa:2019dqn}.
We do not explicitly show $T$ for the argument of the partition function below because we are interested in the $\mu$-dependence with fixed $T$.

With the fugacity expansion, we have the grand canonical partition function expressed as
\begin{align}
    {\cal Z}_\mathrm{GC}(\mu_\mathrm{R})
    &= \sum_{k=-\infty}^\infty
       \exp \Bigl( 3 k \frac{\mu_\mathrm{R}}{T} \Bigr)
       {\cal Z}_\mathrm{C}(3 k)
       \nonumber\\
    &= \sum_{k=-\infty}^\infty
       \exp \Bigl( Q' \frac{\mu_\mathrm{R}}{T} \Bigr)
       {\cal Z}_\mathrm{C}(Q'),
\label{Eq:fugacity_RW}
\end{align}
where $Q' = 3k$ and we use the well known fact that $N_\mathrm{c}$ multiples of $n$ only contribute ${\cal Z}_\mathrm{C}$ because of RW periodicity, as shown in Eq.\,(\ref{Eq:canonical_RW_p}); see also Ref.\,\cite{Roberge:1986mm}.
In the above, we start from ${\cal Z}_\mathrm{GC}(\theta)$ to obtain ${\cal Z}_\mathrm{C}(Q)$, but we can reverse it.

At high $T$, physical quantities such as partition function, entropy, density, etc., should oscillate gently and smoothly at finite $\theta$ on a particular replica; the RW transition, which is a singularity, is induced by the transfer from one replica to another replica.
Thus, we can assume the form of ${\cal Z}_\mathrm{GC}(\theta)$ as
\begin{align}
    {\cal Z}_\mathrm{GC}(T,\mu_\mathrm{I}) &= \frac{1}{2}a_0 + \sum_{\ell=1}^{\ell_\mathrm{max}} a_\ell \cos(\ell \theta),
\label{eq:fd}
\end{align}
for one particular $\mathbb{Z}_{N_\mathrm{c}}$ image, where $\ell \in \mathbb{Z}$ and $\ell_\mathrm{max}$ denote the cutoff of the sum, which is essentially $+\infty$, and $a_\ell \in \mathbb{R}$ refers to the Fourier coefficients; see Ref.\,\cite{Kashiwa:2017swa} for an actual fitting example.
This is noting but the Fourier series.
If the form of the oscillation behavior is gentle and smooth, $m_\mathrm{max}$ can be set as a small number.
This functional form only has the trivial $2\pi$ periodicity, but the RW periodicity can be reproduced by considering its $\mathbb{Z}_{N_\mathrm{c}}$ replicas.
After performing the integration in Eq.\,(\ref{Eq:canonical}), only the $\ell=Q$ terms survive in the canonical partition function due to the properties of the Fourier decomposition.
Therefore, taking into account the properties of the RW periodicity\,(\ref{Eq:canonical_RW_p}), we obtain the grand canonical partition functions as
\begin{align}
    {\cal Z}_\mathrm{GC}(\mu_\mathrm{R})
    &= \frac{1}{2} \sum_{k=-\ell_\mathrm{max}}^{\ell_\mathrm{max}}
       \exp \Bigl( {3k \frac{\mu_\mathrm{R}}{T}} \Bigr)
       a_{3 k}
    \nonumber\\
    &= \frac{1}{2} \sum_{k=-\ell_\mathrm{max}}^{\ell_\mathrm{max}}
       \exp \Bigl( {Q' \frac{\mu_\mathrm{R}}{T}} \Bigr)
       a_{Q'}
    .
\label{Eq:Zc_RW}
\end{align}
We can see that $Q'=3k$ contributions only survive.

In contrast, if we consider the analytic continuation or the Taylor expansion, the non-$3k$ contributions can exist.
The effective potential (the grand potential), expressed using the Taylor expansion, is given as
\begin{align}
    \Omega(T,\mu_\mathrm{R})
    &= c_0 + c_2 \mu_\mathrm{R}^2 + c_4 \mu_\mathrm{R}^4 + \cdots, \\
    \Omega (T,\mu_\mathrm{I})
    &= c_0 - c_2 \mu_\mathrm{I}^2 + c_4 \mu_\mathrm{I}^4 + \cdots, 
    \label{eq:Taylor}
\end{align}
where $c_n \in \mathbb{R}$ are Taylor coefficients.
If there are no singularities at $\mu^2 = (0,0)$, we can use the analytic continuation from the imaginary to the real chemical potential regions until a phase transition occurs.
The Taylor coefficients in Eqs.\,(\ref{eq:Taylor}) are the same at finite $\mu_\mathrm{R}$ and $\mu_\mathrm{I}$, of course.
This means that several quantities have the $Q \neq 3k$ contributions if the Taylor coefficients contain them at finite $\mu_\mathrm{I}$.
To discuss the Taylor coefficients model independently, we consider the perturbative one-loop calculation and the calculation with the strong coupling expansion in Sec.\,\ref{sec:ht} and \ref{sec:sc}, respectively.

\subsection{High temperature region above RW endpoint}
\label{sec:ht}

The perturbative one-loop effective potential with the semi-classical expansion of the gluon field  and the Polyakov-gauge fixing for the quark and gluon contributions is expressed as
\begin{align}
    \Omega(T,\mu_\mathrm{I}) 
    &= \frac{4 N_\mathrm{f} T^2}{\pi^2} 
       \sum_{c=1}^3 \sum_{n=1}^\infty
       \frac{m^2 K_2\Bigl( \frac{n m}{T} \Bigr)}{n^2}
       \cos (2\pi n x),
       \label{eq:ep_p}
\end{align}
where
\begin{align}
    x &= g {\cal A}_4^c + \theta + \pi,
\end{align}
here $g$ is the gauge coupling constant, $m$ represents the current quark mass, $N_\mathrm{f}$ is the number of flavors, $K_2(x)$ denotes the modified Bessel function of the second kind, and ${\cal A}_4^c$ is the diagonal (semi-)classical part of the temporal gluon field $A_4^c$ in the semi-classical expansion scheme with color index $c$, which is defined via the minimization of $\Omega$; for example, see Ref.\,\cite{Sakamoto:2007uy}.
At sufficiently high $T$, $g{\cal A}_4^c$ can take $0$, $\frac{2}{3}\pi$ and $\frac{4}{3}\pi$ depending on the value of $\theta$ because the absolute value of the Polyakov loop must be $1$; they are nothing but the $\mathbb{Z}_3$ images.
In other words, the trivial $\mathbb{Z}_3$ image characterizes the trivial Polyakov-loop sector, and the nontrivial images characterize the nontrivial Polyakov-loop sectors.

In the chiral limit, without spontaneous chiral symmetry breaking, Eq.\,(\ref{eq:ep_p}) becomes
\begin{align}
    \Omega(T,\mu_\mathrm{I})
    &= \frac{4 N_\mathrm{f} \pi^2 T^4}{3} B_4(x)
    \nonumber\\
    &= \frac{4 N_\mathrm{f} \pi^2 T^4}{3}
       \Bigl( x^4 -2x^3 + x^2 -\frac{1}{30}\Bigr), 
\label{eq:ep_p2}
\end{align}
because the asymptotic representation of $K_2(x)$ is known as
\begin{align}
    K_2(x) &= \frac{1}{x^2} - \frac{1}{2} + \cdots,
\end{align}
where $B_4(x)$ denotes the fourth-order Bernoulli polynomial, which has the following relation with the $\cos$ function as
\begin{align}
    B_4 (x) &= -\frac{3}{\pi^4} \sum_{n=1}^\infty \frac{\cos(2 \pi n x)}{n^4}.
\end{align}

We can clearly see that there are $n \neq 3k$ contributions in the region from Eqs.\,(\ref{eq:ep_p}) and (\ref{eq:ep_p2}).
This means that the Taylor coefficients in Eqs.\,(\ref{eq:Taylor}) should obviously have $n \neq 3k$ contributions.
It should be noted that $n \neq 3k$ contributions vanish in the canonical approach because of the RW transition and the RW periodicity, as seen in Eq.\,(\ref{Eq:Zc_RW}).

\subsection{Low temperature region below RW endpoint}
\label{sec:sc}

Qualitative properties of the low temperature region can be inferred from the strong coupling limit because it represents the confined and chiral symmetry broken system; for example, see Refs.\,\cite{Nishida:2003fb,Kawamoto:2005mq} as an example.
The thermodynamic potential in the strong coupling limit is
\begin{align}
    \Omega(T,\mu_\mathrm{I}) & \sim T \ln\Bigl[ \frac{1}{4} \cos(3 \theta) \Bigr].
\end{align}
This only includes the $3k$ contributions, and it is natural from the viewpoint of the RW periodicity without singularities, unlike the high $T$ region.

It should be noted that higher-order terms of the effective potential with the strong coupling expansion contain the following parts in the effective potential;
\begin{align}
       \Phi e^{ i\theta},~~~\Phi e^{-2i\theta},
    ~~~{\bar \Phi} e^{-i\theta},~~~{\bar \Phi} e^{2i\theta},
    \label{eq:parts}
\end{align}
where $\Phi$ denotes the Polyakov loop,
\begin{align}
    \Phi &= \mathrm{tr_c} \Bigl[ {\cal P}  
                 \exp \Bigl( i g \oint d \tau {\cal A}_4^c \Bigr)
                        \Bigr],
\end{align}
here ${\cal P}$ denotes the path ordering, $\mathrm{tr_c}$ means the trace acting on the color space, $\tau$ is the imaginary time, and ${\bar \Phi}$ denotes the conjugated value of the Polyakov loop.
Therefore, it seems that it has non-$3k$ contributions; for example, see Ref.\,\cite{Miura:2016kmd}.
However, they change into $3k$ contributions because the parts (\ref{eq:parts}) are transformed as
\begin{align}
    \Psi,~~~\Psi e^{- 3i\theta},~~~{\bar \Psi},~~~ {\bar \Psi} e^{3i\theta},
\end{align}
where $\Psi$ is the so called modified Polyakov loop
\begin{align}
\Psi \equiv \Phi e^{i\theta}.
\end{align}
The modified Polyakov loop is the RW periodic quantity based on the extended ${\cal Z}_3$ symmetry related to the following transformation~\cite{Sakai:2008py};
\begin{align}
    \Phi &\to \Phi \exp\Bigl( - \frac{2 \pi i k}{3} \Bigr),
    \nonumber\\
    {\bar \Phi} &\to {\bar \Phi} \exp\Bigl( \frac{2 \pi i k}{3} \Bigr),
    \nonumber\\
    e^{\pm i\theta} &\to e^{\pm i\theta} \exp\Bigl( \pm \frac{2 \pi i k}{3} \Bigr).
\end{align}
The modified Polyakov loop is invariant under the extended ${\cal Z}_3$ symmetry, and this symmetry is nothing but the shift symmetry~\cite{Kashiwa:2012xm,kikuchi2018t,Nishimura:2019umw,Shimizu:2017asf}; for example, see Ref.\,~\cite{Kashiwa:2019dqn}.
Above the RW endpoint temperature, the extended ${\cal Z}_3$ symmetry is spontaneously broken, and then non-$3k$ contributions can appear because of the singularities originating from the RW transition.
Below the temperature of the RW endpoint, the symmetry is intact, and thus the RW transition does not occur, and only $3k$ contributions survive.

Based on the above explanations, Taylor coefficients only have $n = 3k$ contributions in the low $T$ region, and thus there is no mismatch with the canonical approach in this region.
Of course, it is trivial because all quantities are smooth and continuous in this region due to the absence of the RW transition, and thus must be described by the $2\pi/3$ periodic functions; this means that there are only $n=3k$ contributions.
It should be noted that there may be a chiral symmetry induced first order transition very close to the RW endpoint, but it does not affect the RW periodicity, and thus it is not important in our present discussion.

\section{Discussions}
\label{sec:Discussions}

At low $T$ below the RW endpoint temperature, the effective potential contains only $3k$ contributions at finite $\theta$ due to the RW periodicity without the RW transition.
At high $T$ above the RW endpoint temperature, the RW periodicity is maintained by the $\mathbb{Z}_3$ images; thus, the thermodynamic potential can have any $n \neq 3k$ contribution, as can be seen from the perturbative one-loop potential in principle. 
This indicates that the canonical method is valid below the RW endpoint temperature but not above it.

The mismatch between the correct result and the result obtained by the canonical method is attributable to the expressive power of the Fourier transformation (\ref{Eq:canonical}), as explained below.
Via the Fourier series  and its inverse transformation used in the canonical approach, we have
\begin{align}
    &{\cal Z}_\mathrm{GC}(\theta)
    \nonumber\\
    &\sim \frac{1}{2\pi}
    \sum_{Q = -Q_\mathrm{max}}^{Q_\mathrm{max}} 
    \Bigl[ \Bigl\{ \int_{-\pi}^\pi e^{iQ\theta} {\cal Z}_\mathrm{GC}(\theta) \, d\theta \Bigl\} e^{-iQ\theta} \Bigr],
\label{eq:FIF}
\end{align}
where $Q_\mathrm{max}$ is the upper limit of the Fourier series because we usually need it for numerical calculations.

If there is no information loss in Eq.\,(\ref{eq:FIF}), we can reproduce ${\cal Z}_\mathrm{GC}(\theta)$ starting from ${\cal Z}_\mathrm{GC}(\theta)$; the right-hand side is equal to the left-hand side in Eq.\,(\ref{eq:FIF}).
This equality is trivial if ${\cal Z}_\mathrm{C}(\theta)$ is a smooth, continuous, and $2\pi/3$-periodic function with the limit $Q_\mathrm{max} \to \infty$ from Dirichlet's theorem.
It should be noted that Eq.\,(\ref{eq:FIF}) converges to the correct result by increasing $Q_\mathrm{max}$ in the case that ${\cal Z}_\mathrm{GC}(\theta)$ has the cusp, unlike in the case of a gap.
If there is a gap, the Gibbs phenomenon~\cite{gibbs1898fourier} occurs, and the validity of Eq.\,(\ref{eq:FIF}) is not evident with finite $Q_\mathrm{max}$; there is oscillating overshoot behavior near the gap, and it does not converge to the correct value with increasing $Q_\mathrm{max}$.
With the limit $Q_\mathrm{max} \to \infty$, the series approaches the average of the left-hand and right-hand limits of the gap.
In the case of the cusp, we can reproduce it using the Fourier series and inverse transformation with $Q_\mathrm{max} \to \infty$ due to several conditions; e.g., the Dirichlet condition.

In the Taylor expansion method, we should expand the functions at $\mu=0$ and match with the result of the analytic continuation from $\mu_\mathrm{I}$ and $\mu_\mathrm{R}$.
This means that the Taylor coefficients only have information on the trivial Polyakov loop sector at high $T$.
This means that we should consider only the trivial Polyakov-loop sector also in the canonical approach.
If it is correct, the Fourier series in Eq.\,(\ref{Eq:canonical}) is modified as
\begin{align}
    {\cal Z}_\mathrm{C}(Q) &= \frac{1}{2\pi}
    \Bigl[ \int_{-\pi}^{\pi} e^{iQ\theta} {\cal Z}^\mathrm{fix}_\mathrm{GC}(\theta) \, d\theta \Bigr],
\label{eq:FIF2}
\end{align}
where $ {\cal Z}^\mathrm{fix}_\mathrm{GC}(\theta)$ is the grand canonical partition function that fixes the trivial Polyakov-loop sector; this is the Fourier series of period $2\pi$.
This fixing is, of course, difficult in general, but it is easy in the perturbative one-loop calculation because we just set $g{\cal A}^c_4=0$.
Although the original version (\ref{eq:FIF}) cannot reproduce the result of the Taylor expansion at high $T$ above $T_\mathrm{RW}$ because the original ${\cal Z}_\mathrm{GC}(\theta)$ contains $n \neq 3k$ contributions, but Eq.\,(\ref{eq:FIF2}) can reproduce the correct result.
This fact can be easily seen by using the Fourier decomposition (\ref{eq:fd}) for the trivial Polyakov-loop sector with $m_\mathrm{max} \to \infty$ as
\begin{align}
    & \frac{1}{2\pi}
      \Bigl[ \int_{-\pi}^{\pi} e^{iQ\theta} \Bigl\{\frac{1}{2}a_0 + \sum_{m=1}^\infty a_m \cos(m \theta) 
      \Bigl\}\, d\theta \Bigr]
    \nonumber\\
    &=
    \begin{dcases}
    a_0 ~~~~~~~~~~~(m = 0),
    \\
    \frac{1}{2} a_Q \delta_{Qm}~~~(m \neq 0),
    \end{dcases}
\end{align}
and thus we have
\begin{align}
    {\cal Z}_\mathrm{GC}(T,\mu_\mathrm{R})
    &= \sum_{Q=-\infty}^\infty \xi^Q \Bigl( \frac{1}{2} a_Q \Bigr)
    \nonumber\\
    &\underset{\mu_\mathrm{R}=0}{\longrightarrow}
       \frac{1}{2} a_0 + \sum_{Q=1}^\infty a_Q.
\end{align}
This result matches the correct result (\ref{eq:fd}) with $\theta=0$.
The above facts are the origin of the paradox discussed in this paper.
It should be noted that the original ${\cal Z}_\mathrm{GC}(\theta)$ is a continuous $2\pi/3$-periodic function at low $T$ below $T_\mathrm{RW}$, and therefore naturally does not have $n \neq 3k$ contributions.
Therefore, both Eq.\,(\ref{Eq:canonical}) and Eq.\,(\ref{eq:FIF2}) lead to the same results in the region.
Readers may be concerned about the Lee-Yang zero analysis~\cite{yang1952statistical,lee1952statistical} for QCD because fixing the Polyakov-loop sector seems to affect it.
Actually, the Lee-Yang zero analysis for QCD needs some modifications, as explained in Appendix~\ref{sec:LY}. 

In the lattice QCD simulation, we cannot take an infinite volume as the starting point, and thus the RW transition disappears; we can extract information about the RW transition via extrapolation, as is standard in lattice QCD simulations.
This means that the canonical approach is valid for the lattice QCD simulation, which treats the finite volume system.
However, we may obtain an incorrect result when we take the infinite volume as the starting point for the calculation if we do not fix the Polyakov loop to the trivial sector.
In such a case, we may slightly smear the RW transition by hand if we use the canonical approach.
However, this is not to say that there is no solution in the infinite-size system; it may be possible by using the coarse-graining by the gradient flow~\cite{Narayanan:2006rf,Luscher:2009eq,Luscher:2010iy} because it makes the distribution of the non-averaged Polyakov loops meet up at each $\mathcal{Z}_3$ domain in the deconfined phase after flowing the quantity; it is demonstrated by lattice simulations for the $3$-dimensional Potts model and the quenched QCD~\cite{Ejiri:2026ijj}.
Then, we may obtain the trivial $\mathcal{Z}_3$ image via this procedure even in the infinite-size system.

\section{Summary}
\label{sec:Summary}

In this study, we have revisited the canonical approach at high $T$, particularly in the region above the Roberge-Weiss (RW) endpoint temperature ($T_\mathrm{RW}$) at finite imaginary chemical potential ($\mu_\mathrm{I}$).
We have discussed the validity of the canonical approach when the RW transition exists.

In the low temperature region ($T<T_\mathrm{RW}$), where the RW transition does not exist at finite $\theta=\mu_\mathrm{I}/T$, all $\theta$-even quantities, such as the chiral condensate and thermodynamic quantities, are smooth and continuous functions, and thus the canonical approach is valid; the expressive power of the Fourier series with $2\pi$ period and inverse transformation in the canonical approach is sufficient in this case.
In the high temperature region ($T>T_\mathrm{RW}$), where the RW transition exists at finite $\theta$ in the infinite system, several quantities calculated using the canonical approach cannot be matched with the correct results.
This mismatch originates from the properties of the Fourier series.

The period of the Fourier series used in the canonical approach contains both the trivial and non-trivial Polyakov-loop sectors, and thus the cancellation occurs; all contributions vanish except for those that are multiples of three in the canonical approach.
If quantities only have $3k$ contributions with $k \in \mathbb{Z}$, which are multiples of three quark number contributions, the usual canonical approach can provide correct results.
However, if non-$3k$ contributions survive, we should fix the gluon field in the trivial Polyakov-loop sector.
The cancellation between the trivial and non-trivial sectors is the origin of the paradox discussed in this paper.

When we consider the finite-size system, the  RW transition disappears, and thus $3k$ contributions survive only because the RW periodicity is maintained by the smooth and continuous partition function.
This means the following: The canonical approach is valid over the entire temperature range in the finite size system, but it is invalid in the high temperature region of the infinite size system.
Therefore, we need to exercise extreme care in considering QCD effective models because we usually take the infinite volume as the starting point of the calculation, unlike lattice QCD simulations.
Of course, the canonical method is valid even with the infinite-size system at low $T$ because there is no RW transition.

It should be noted that there is another origin for the breakdown of the canonical approach.
If we use the Feynman diagrammatic approach without employing the semiclassical expansion of the gluon field, non-$3k$ contribution appears even in the low $T$ region because the RW periodicity and the transition have vanished~\cite{Kashiwa:2024cjn}.
Therefore, we need to exercise extreme care when using the canonical approach with the QCD effective models.

\begin{acknowledgments}
This work is supported in part by Grants-in-Aid for Scientific Research from the Japan Society for the
Promotion of Science (JSPS) KAKENHI (Grant No. JP22H05112).
\end{acknowledgments}

\appendix

\section{Lee-Yang zeros}
\label{sec:LY}

If we adopt the Fourier series shown in Eq.\,(\ref{eq:FIF2}), we can partially reproduce the RW transition from the point of view of the Lee-Yang zero analysis.
The following discussions are almost the same as those in Ref.\,\cite{Nagata:2014fra}, but the effective potential has the $2\pi$ period because we fix the Polyakov-loop sector to the trivial one.

In the high-$T$ region, the grand canonical partition function can be expressed as
\begin{align}
    {\cal Z}_\mathrm{GC}(\mu)
    &\sim \sum_{k=-\infty}^\infty \frac{1}{2\pi}
       \exp \Bigl( - \frac{n}{4T^3 V_3 c_2} + \frac{\mu}{T}\Bigr),
\label{eq:GCP_2}
\end{align}
where $\mu$ is the complex chemical potential, $c_2$ is the coefficient of $2$-nd order in the Taylor expansion of the effective potential, and $V_3$ denotes the volume of the three-dimensional system.
Equation (\ref{eq:GCP_2}) is nothing but the theta function defined as
\begin{align}
    \vartheta(z,\tau) &= \sum_{k=-\infty}^\infty \exp \Bigl( \pi i k^2 \tau + 2 \pi i k z\Bigr),
\end{align}
where we have the following correspondences with Eq.\,(\ref{eq:GCP_2});
\begin{align}
    2 \pi i z &= \frac{\mu}{T},~~~\pi i \tau = - \frac{1}{4 T^3 V_3 c_2}.
\end{align}
It is known that the theta function has zeros as
\begin{align}
    z &= \frac{2 k+1}{2} + \frac{2l+1}{2} \tau,
\end{align}
where $k,~l \in \mathbb{Z}$.
Therefore, Eq.\,(\ref{eq:GCP_2}) has zeros at
\begin{align}
    \frac{\mu}{T} &= (2k+1) \pi i - \frac{2l-1}{4 T^3 V_3 c_2}, 
\end{align}
and thus we can read off the transition at $\theta=\pi$ in the limit $V_3 \to \infty$.
This means that the RW transition point is partially reproduced if we adapt the fixing of the trivial Polyakov-loop sector.
To reproduce all other RW transition points, $\theta= \pi/3$ and $5\pi/3$, we should fix the Polyakov-loop sector to the non-trivial sectors. 

If we consider the finite size system, the grand canonical partition function naturally becomes a smooth and continuous $2\pi/3$ periodic function against $\theta$.
In this case, there is no need to fix the calculation to the trivial Polyakov-loop sector because there is no paradox there, as discussed in the main text.

\bibliography{ref.bib}
\end{document}